# Quantum key distribution using different-frequency photons

B.-S. Shi*, Y.-K. Jiang, G.-C. Guo**

Lab. of Quantum Communication and Quantum Computation and Department of Physics, University of Science snd Technology of China, Hefei, 230026, P.R. China

**Abstract.** In this paper, we present a scheme for quantum key distribution, in which different-frequency photons are used to encode the key. These different-frequency photons are produced by an acoustic-optical modulator and two kinds of narrow filters. This scheme may be implementable in practice.

Quantum cryptography allows two remote parties to exchange information secretly by a sequence of secret quantum keys. The quantum key distribution provides a means to establish such a key at a distance and to check its confidentiality, its security is based on the fact that any measurement of incompatible quantities on a quantum system by an eavesdropper will inevitably modify the state of this system. Therefore, an eavesdropper might get some or full information out of a quantum channel, but will leave some traces, which can be detected by legal users. The idea of quantum cryptography germinated in 1970 [1], the first quantum key distribution model was present by Bennett and Brassard in 1984 [2], and the first key distribution experiment was completed in 1992 [3]. Since then, a lot of theoretical models and practical methods have been reported [4–22].

In practice, a single photon is used as an information carrier, the key can be encoded by its polarization or its phase. In this paper, we present a cryptosystem, in which different-frequency photons are used to encode the key. In this system, an acoustic-optical modulator (AOM) is used to generate different-frequency photons. In the sender's kit, two kinds of narrow filters are used to encode the key by sender (Alice), and in the receiver's kit, receiver (Bob) receives the information sent from Alice by switching on or off the AOM randomly in each case. We design this cryptosystem based on an idea [17, 20], such that a "plug and play" system as in [17, 20] may be set up. One problem of this system is how to control the narrow filter with high speed, it may affect the implementation of this system in practice.

If quantum key distribution is to become widespread, it should be effective over an optical network. Townsend et al. have shown how the properties of a passive optical network can be exploited to give one-to-any key distribution on branch-and loop-configuration networks [21], and Phoenix et al. have presented an any-to-any key distribution model on optical networks [22]. In this paper, we show how this scheme can be extended to distribute a quantum key between sender and any one of $N$ receivers on a branch-configuration network. In this scheme, only one AOM is used. It is may be implementable in practice.

In our scheme, three different frequency photons are used, which are $|\omega\rangle$, $|\omega+\delta\rangle$, and $1/\sqrt{2}[\,|\omega\rangle + |\omega+\delta\rangle]$, respectively, and $\langle\omega|\omega+\delta\rangle = 0$, $\langle\omega|1/\sqrt{2}[|\omega\rangle + |\omega+\delta\rangle] = 1/\sqrt{2}$. The frequency states $|\omega\rangle$ and $|\omega+\delta\rangle$ are used to encode the key, which stand for binary "0" and "1", respectively. The frequency state $1/\sqrt{2}[|\omega\rangle + |\omega+\delta\rangle]$ is used as a control state, which is used to detect whether an eavesdropper (Eve) is present or not. These three states can be produced by an AOM and two kinds of narrow filters. The setup of our scheme is shown in Fig. 1.

Bob initiates the transmission by sending a short pulse produced by source (S) to Alice. In Bob's kit, a monochromatic beam of frequency $\omega$ is introduced into an AOM driven at radio frequency (rf) $\delta$. The incident wave is separated into two equal-amplitude beams, one is the transmitted beam and the other the diffracted beam. The frequencies of the transmitted and diffracted waves are $\omega$ and $\omega+\delta$, respectively. The diffracted wave is delayed and the delay time is greater than the coherence time of pulses. A polarization controller is set so that the wave $\omega$ is completely transmitted and the wave $\omega + \delta$ is reflected completely at the polarizing beam-splitter (PBS). These two beam are introduced into a long-distance channel connecting with Alice. For encoding the key, Alice selects randomly one of two narrow filters $F(\omega)$ or $F(\omega + \delta)$ and inserts into channel or lets the channel free-filter, then the wave is reflected by a Faraday mirror (FM) back to Bob. Thanks to the FM, the birefringence of the optical link is compensated, and the pulse comes back orthogonally polarized. In Bob's kit, the wave $\omega$ is reflected by the PBS into the long arm, and the wave $\omega + \delta$ is transmitted into the short arm. The detection on Bob's side is done by switching the AOM with rf $\delta$ on or off randomly. First, we discuss the case in which Bob switches on the AOM. If there is no filter in the channel in Alice's side, because of constructive interference, only the wave $\omega$ in beam 1 can be detected by detector $D_1$. The interference results from the superposition of two undistinguished paths: AOM → $\omega$ → short→PBS→FM→PBS→ long→AOM and AOM → $\omega + \delta$ → long→PBS→FM→PBS→ short→AOM. If Alice inserts one of two kinds of filters, the interference is destroyed, so there is 50% probability that the detector $D_1$ or $D_2$ fires, respectively. In this case, Bob can not distinguish the states sent by Alice at all. Secondly, we discuss the case in which Bob switches off the AOM. In this case, the AOM can be regarded as a transparency media. If there is no filter in Alice's kit, the detector $D_1$ and detector $D_2$ can fire with a probability of 50%, respectively, Bob can not get any information about the state sent by Alice too. If Alice inserts a narrow filter $F(\omega)$, the wave $\omega$ will transmit through the AOM and reach the detector $D_1$, detector $D_2$ is dark, this will happen with 50% probability. If the filter $F(\omega + \delta)$ is inserted, then the wave $\omega + \delta$ will reach the detector $D_2$, the detector $D_1$ can not fire, the probability is also 50%. In these two cases, Bob can know exactly the state from Alice.

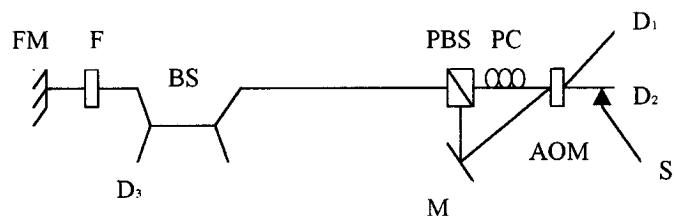

**Fig. 1.** One-to-one quantum key distribution scheme

*E-mail: drshi@ustc.edu.cn
**E-mail: gcguo@ustc.edu.cn

In order to distribute the quantum key, prevent a malevolent eavesdropper called Eve to divert part of the pulse to get information about the key, the single-photon source should be used (but this is unpractical, the very weak pulse of light, about 0.1 photon per pulse on average is used instead of a single photon). We should remember that the pulses going from Bob to Alice do not carry any information, it is only on the way back to Bob that the information is encoded by filters by Alice. In order to detect any attempt to obtain the value of the frequency by sending a much stronger pulse in the system and measuring the frequency of the reflected pulses by diverting part of pulse, a 50/50 beam splitter (BS) and a single-photon detector $D_3$ is used to monitor the intensity of the pulses from Bob as in [22]. Besides, because the transmission rate of the filter is relatively low, when the filter in Alice's kit is removed, the intensity of the reflected light will increase obviously. To avoid an eavesdropper detecting this change, a fast attenuator should be used. This may be not convenient in practice.[1]

Now, we discuss how to distribute the key between Alice and Bob. To encode the key, Alice selects randomly the filter $F(\omega)$ or $F(\omega + \delta)$ or nothing to insert into the channel. Bob receives the information from Alice by switching on or off the AOM randomly in each case. After having transmitted a sequence of photons, Alice and Bob decide to discard all instances in which none of two detectors fires, and divert the remaining instances into three groups by classical communication. One group consists of instances in which Alice inserts nothing into the channel and Bob switches on the AOM simultaneously. In this group, only detector $D_1$ fires because of constructive interference if no eavesdropping is present during transmission. This group can be used to detect whether an eavesdropper is present or not. Because if someone eavesdrops during the transmission, the interference is destroyed and there is 50% probability that detector $D_2$ will fire instead of $D_1$. The second group consists of instances in which Alice chooses one of two filters to insert into the channel and Bob switches off the AOM at the same time. In this group, if the detector $D_1$ fires, Bob can deduce that the state sent by Alice is $|\omega\rangle$. If $D_2$ fires, he knows exactly that the state $|\omega + \delta\rangle$ is sent by Alice. According to coding principle, these states are changed into binary bit "0" or "1", respectively. The third group consists of instances in which Alice chooses one of two filters and Bob switches on the AOM. This group will be discarded because Bob can not distinguish exactly the state sent by Alice.

Our scheme may be designed to realize a "plug and play" system as in [17, 20]. Our system faces a difficulty: how to insert or remove filter in Alice's kit with high speed? It will have an effect on the implementation of this scheme in practice. One possible way is: the selection of filters can be realized by optical switching among three paths in which, two of the three paths have filters and the other has nothing.

In the following, we show how to give one-to-any key distribution using this system on a branch-configuration network. Our branch-configuration network is shown in Fig. 2, which is based on the idea of [23]. A broadcaster, Alice, sends a sequence of single photons into a passive optical network. The law of quantum mechanics dictates that single quanta, such as photons, can neither be split nor cloned [24]. A given single photon can only reach a single receiver, Bob($n$), so an individual key between Alice and each one of Bobs can be established separately using our system. There is only one AOM in this scheme, it may be realized in practice. This network has a characteristic that only Alice knows which Bob receives the signal in each instance, because she knows the time of transmission and range between her and each Bob, so the ranges between Alice and each Bob should be different if this network works.

In summary, we present a proposal for quantum key distribution using different-frequency photons, in which, one AOM and two narrow filters are used to produce different-frequency photons, and encode the key. We also extend this scheme into one-to-any network on branch-configuration, in which only one AOM is used. These two schemes may be implementable in practice.

---

[1] We thank the referee for pointing out this important point.

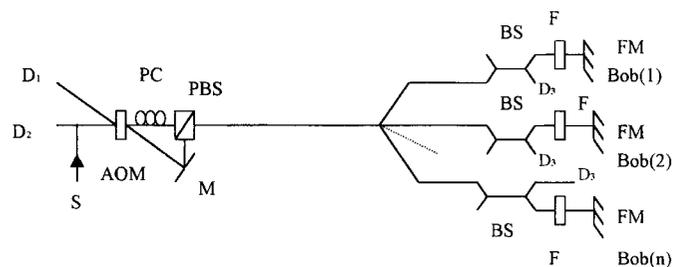

Fig. 2. One-to-any quantum key distribution on branch-configuration

*Acknowledgements.* We are grateful to the referee for helpful remarks and stimulating suggestions. This subject is supported by the National Natural Science Foundation of China.